%% file: main.tex
\documentclass[10pt,conference]{IEEEtran}
\IEEEoverridecommandlockouts
\usepackage[T1]{fontenc}
\usepackage{amsmath,amssymb}
\usepackage{cite}
\let\citep\cite
\usepackage{url}
\usepackage[hidelinks]{hyperref}
\hypersetup{pdftitle={Rethinking the Value of Agent-Generated Tests for LLM-Based Software Engineering Agents},pdfauthor={Zhi Chen; Zhensu Sun; Yuling Shi; Chao Peng; Xiaodong Gu; David Lo; Lingxiao Jiang}}
\providecommand{\Description}[1]{}
\usepackage{booktabs}
\usepackage{tabularx}
\usepackage{multirow}
\usepackage{array}
\usepackage{longtable}
\usepackage{listings}
\usepackage{graphicx} 
\usepackage{makecell}
\usepackage{bbm}
\usepackage{adjustbox}
\usepackage{threeparttable}
\usepackage{balance}
\usepackage[table]{xcolor}
\usepackage{enumitem}
\usepackage{placeins}
\usepackage{tikz}
\usetikzlibrary{arrows.meta,positioning}
\usepackage{tcolorbox}
\tcbuselibrary{skins,breakable}
\definecolor{takeawaygreen}{RGB}{138,179,137}
\newtcolorbox{takeawaybox}[1][]{
  enhanced,
  attach boxed title to top left={xshift=4mm,yshift=-2mm},
  colback=takeawaygreen!10,
  colframe=takeawaygreen!60,
  colbacktitle=takeawaygreen!80,
  coltitle=black!85,
  fonttitle=\bfseries\small,
  fontupper=\small,
  boxed title style={size=small,colframe=takeawaygreen!80,sharp corners},
  sharp corners,
  boxrule=0.8pt,
  left=4pt,right=4pt,top=10pt,bottom=4pt,
  before skip=7pt,after skip=7pt,
  unbreakable,
  title={#1}
}

\begin{document}
\bstctlcite{bibstylecontrol}
\title{Rethinking the Value of Agent-Generated Tests for LLM-Based Software Engineering Agents}
\author{
\IEEEauthorblockN{\makebox[0.31\textwidth]{Zhi Chen}}
\IEEEauthorblockA{Singapore Management University\\
Singapore, Singapore\\
zhi.chen.2023@smu.edu.sg}
\and
\IEEEauthorblockN{\makebox[0.31\textwidth]{Zhensu Sun\textsuperscript{*}\thanks{\textsuperscript{*}Corresponding author.}}}
\IEEEauthorblockA{Singapore Management University\\
Singapore, Singapore\\
zssun@smu.edu.sg}
\and
\IEEEauthorblockN{\makebox[0.31\textwidth]{Yuling Shi}}
\IEEEauthorblockA{Shanghai Jiao Tong University\\
Shanghai, China\\
yuling.shi@sjtu.edu.cn}
\and[\hfill\mbox{}\par\mbox{}\hfill]
\IEEEauthorblockN{\makebox[0.31\textwidth]{Chao Peng}}
\IEEEauthorblockA{The University of Edinburgh\\
Edinburgh, United Kingdom\\
chao.peng@acm.org}
\and
\IEEEauthorblockN{\makebox[0.31\textwidth]{Xiaodong Gu}}
\IEEEauthorblockA{Shanghai Jiao Tong University\\
Shanghai, China\\
xiaodong.gu@sjtu.edu.cn}
\and
\IEEEauthorblockN{\makebox[0.31\textwidth]{David Lo}}
\IEEEauthorblockA{Singapore Management University\\
Singapore, Singapore\\
davidlo@smu.edu.sg}
\and[\hfill\mbox{}\par\mbox{}\hfill]
\IEEEauthorblockN{\makebox[0.31\textwidth]{Lingxiao Jiang}}
\IEEEauthorblockA{Singapore Management University\\
Singapore, Singapore\\
lxjiang@smu.edu.sg}
}
\maketitle
\input{sections/abstract}
\input{sections/introduction}
\input{sections/methodology}
\input{sections/RQ1}
\input{sections/RQ2}
\input{sections/RQ3}
\input{sections/discussion}
\input{sections/related_work}
\input{sections/conclusion}
\input{sections/data_availability}
\IEEEtriggercmd{\balance}
\IEEEtriggeratref{43}
\bibliographystyle{IEEEtran}
\bibliography{bib-style,sample-base}
\end{document}

%% file: sections/abstract.tex
\begin{abstract}
Large Language Model (LLM) code agents increasingly resolve repository-level issues by iteratively editing code, invoking tools, and validating candidate patches. In these workflows, agents often write tests on the fly, but the value of this behavior remains unclear. This raises a central question: how does agent-written testing relate to task outcomes and resource use during repository-level issue resolution?

To better understand the role of agent-written tests, we analyze 5{,}000 recorded agent runs produced by six strong LLMs on SWE-bench Verified using the lightweight \texttt{mini-SWE-agent} scaffold.
Our results show that test writing is common for most models. GPT-5.2 is an exception: it rarely writes new test files while achieving a resolution rate close to the top-ranked model. Test-writing frequency varies widely across models. Within a given model, it differs less between resolved and unresolved tasks. When tests are written, their files contain substantially more statements that print runtime values than assertions that check conditions. We then compare baseline prompts with revised prompts that encourage or discourage writing new test files for four models. Agents start or stop writing test files as requested on 36.8\%--93.2\% of tasks. Resolved-task totals differ by 0--13 out of 500 per model, with no statistically significant difference, while API calls or token use differ substantially for three models. Similar resolution totals can therefore hide substantial differences in how agents test and how many resources they use.
\end{abstract}

%% file: sections/introduction.tex
\section{Introduction}
\label{sec:intro}
Code agents combine Large Language Models (LLMs)~\cite{openai_gpt52_systemcard_2025,chen2024promise} with tools to inspect repositories, edit code, and resolve software issues~\cite{yang2024sweagent,wang2024openhands,hong2023metagpt,ehrlich2025codemonkeys,xia2024agentless}. We use \emph{code agent} to denote an LLM operating in an iterative action--observation loop, and \emph{scaffold} to denote the interface and protocol that govern its tools and feedback. Testing supports this workflow by exposing regressions and checking hypotheses during patch development~\cite{zhang2024autocoderover,xia2024agentless,liu2024marscode,mu2025experepair}.

Repository-level agents can validate patches with existing human-written tests, which reflect project conventions and developer intent~\cite{jimenez2024swebench,chen2024evaluating}, or with \emph{agent-written tests} created during issue resolution. Agent-written tests can surface edge cases and support fault localization, but incorrect expectations about program behavior can redirect the repair process. Writing and running them also consumes API calls, tokens, and context~\cite{kim2025towards}.
These temporary tests give agents a way to inspect program behavior through execution and printed values, connecting testing with program comprehension. Resolution rate alone does not capture this activity or its resource use.

To better understand agent-written tests, we conduct a quantitative analysis of agent trajectories on SWE-bench Verified~\cite{swebench_verified} using \texttt{mini-SWE-agent}~\cite{mini_swe_agent}. Its default prompt recommends testing edge cases and leaves whether, when, and how to test to the model. Agent-written testing is prevalent for several strong models in this setup.
For example, \textit{Claude Opus 4.5} (ranked \#1 in this setting, 74.4\% resolution) writes a detected test file in 83.6\% of tasks.
By contrast, \textit{GPT-5.2} achieves a comparable resolution rate (71.8\%), only 2.6 percentage points below \textit{Claude Opus 4.5}, while writing detected test files in only 2.4\% of tasks.
This observation motivates a core question: \emph{how does agent-written testing relate to task outcomes and resource use during repository-level issue resolution?}
Answering this question requires a systematic empirical study of testing behavior, test-file content, and differences across test-writing prompt conditions.

Prior work mostly evaluates LLM-generated unit tests, assertions, or issue-reproducing tests against a fixed target program or code snapshot~\cite{lops2025system,zhang2025exploring,mundler2024swt,yuan2024evaluating,wang2025testeval,wang2024software,schafer2023empirical}. However, in complex real-world GitHub issue resolution~\cite{jimenez2024swebench,khandpur2025swebenchmultilingual}, the codebase and candidate patches evolve over time, and testing occurs within the broader repair process rather than as a standalone evaluation objective. Existing studies commonly focus on either generated test artifacts or agent trajectories. We connect these perspectives by examining how strong models write and execute tests within repository-level repair under a lightweight scaffold and how outcomes and resource use differ across test-writing prompts.

\paragraph{Research questions and overview}
We organize the study around three research questions.
\textbf{RQ1} characterizes the agents' testing behaviors under a lightweight scaffold where test writing is optional: whether agents write tests, when they introduce them, and how intensively they execute them.
\textbf{RQ2} shifts from behaviors to \emph{test content}, investigating how many assertions and value-revealing prints appear in agent-written test files, what forms the assertions take, and what kinds of runtime information the prints expose.
\textbf{RQ3} compares paired prompt conditions that encourage or discourage writing new test files, measuring differences in test-writing behavior, task resolution, API calls, and tokens.

The contributions of this work can be summarized as follows:
\begin{itemize}[leftmargin=*]

\item \textbf{A cross-model analysis of agent-written testing behavior.}
Across 3{,}000 baseline trajectories, we characterize whether, when, and how six LLMs create and execute test files, finding wide behavioral variation that does not follow the ranking of resolution rates.

\item \textbf{An analysis of assertions and value-revealing prints in agent-written test files.}
Rule-based analyses of Python abstract syntax trees (ASTs) characterize the conditions that assertions check and the runtime information that prints expose. Across the five analyzed models, statements that display runtime values outnumber assertions.

\item \textbf{A comparison of task outcomes and resource use across test-writing prompts.}
Across 2{,}000 task-level pairs, agents start or stop writing test files as requested on 36.8\%--93.2\% of tasks; none of the four model-level outcome differences is statistically significant, while resource use differs substantially for three models.

\end{itemize}

%% file: sections/methodology.tex
\section{Methodology} \label{sec:methodology}
Figure~\ref{fig:study-design} summarizes the study design. We analyze 3{,}000 baseline trajectories from six models on the same 500 tasks for RQ1. RQ2 examines test-file content from the five models that produce enough test files for comparison. For RQ3, we run four models with revised prompts and pair each run with the same model's baseline on the same task. These 2{,}000 additional runs give 2{,}000 baseline--revised pairs and bring the study total to 5{,}000 agent trajectories.

\begin{figure*}[t]
  \centering
  \input{images/study_design_overview}
  \caption{Across 5{,}000 \texttt{mini-SWE-agent} trajectories on 500 SWE-bench Verified tasks, RQ1 characterizes testing behavior, RQ2 examines test-file content, and RQ3 compares baseline runs with prompts that encourage or discourage new test writing. The analyses use 3{,}000 baselines for RQ1, a five-model content subset for RQ2, and 2{,}000 baseline--revised task pairs for RQ3.}
  \Description{The study uses 500 SWE-bench Verified tasks and the mini-SWE-agent scaffold. The baseline branch contains 3,000 trajectories from six models and supports RQ1 and RQ2. The revised-prompt branch adds 2,000 trajectories from four models, paired with their baselines, and supports RQ3.}
  \label{fig:study-design}
\end{figure*}

\subsection{Benchmark} \label{sec:methodology-models}

We use SWE-bench Verified as our benchmark. The original SWE-bench benchmark is built from resolved GitHub issues drawn from 12 open-source Python repositories~\cite{jimenez2024swebench}. SWE-bench Verified is a 500-instance subset released after a human-screening effort led by OpenAI in collaboration with the SWE-bench authors~\cite{swebench_verified}. Each instance provides a GitHub issue, a fixed repository snapshot, and an evaluation harness that checks whether a patch passes the issue-specific and regression tests. We use this official harness outcome to label each task as resolved or unresolved.

\subsection{Agent Scaffold and Models}
Many recent LLM-based agents incorporate dedicated testing components, such as specialized validation modules, test-planning stages, or multi-agent coordination~\cite{liu2024marscode,zhang2024autocoderover,ruan2024specrover}.
We adopt \texttt{mini-SWE-agent}~\cite{mini_swe_agent} as a lightweight scaffold in which testing choices remain model-controlled.
The agent interacts with the repository through a single \texttt{bash} tool, using shell commands to inspect and modify files and run programs such as Python.
The model can create executable Python test files on the fly and run them via \texttt{bash} as part of its workflow.
The default \texttt{mini-SWE-agent} prompt includes a brief natural-language recommendation: ``Test edge cases to ensure your fix is robust.''
Beyond this brief cue, the model controls whether, when, and how to test through the standard \texttt{bash} interface.
Our measurements characterize how each model tests in this prompt-and-scaffold setup.

We download the publicly released baseline trajectories from the SWE-bench \emph{Bash Only} leaderboard\footnote{\url{https://www.swebench.com/}}.
For model selection, we use the SWE-bench \emph{Bash Only} leaderboard with a cutoff date of 2025-12-11 and identify the top-six \emph{model families} (rather than top entries, since a family may appear with multiple variants on the leaderboard).
For each family, we use its highest-ranked model as the representative:
\textit{claude-opus-4.5}~\cite{anthropic_claude_opus45_2025} (74.4\%),
\textit{gemini-3-pro-preview}\footnote{Google's Gemini 3 announcement: \url{https://blog.google/products/gemini/gemini-3/}} (74.2\%),
\textit{gpt-5.2}~\cite{openai_gpt52_systemcard_2025} (71.8\%),
\textit{kimi-k2-thinking}~\cite{moonshot_kimi_k2_thinking_2025} (63.4\%),
\textit{minimax-m2}~\cite{minimax_m2_2025} (61.0\%),
and \textit{deepseek-v3.2-reasoner}~\cite{deepseek_v32_2025} (60.0\%).
In the remainder of the paper, including tables and figures, we refer to these models as Claude Opus 4.5, Gemini 3 Pro, GPT-5.2, Kimi K2 Thinking, MiniMax M2, and DeepSeek v3.2 Reasoner.
\subsection{Data Extraction of Agent-Written Tests}
\label{sec:methodology-collection}

We extract agent-written test files from trajectories: time-ordered logs of the agent's reasoning, bash actions, and environment observations. We call each detected Python test file a \emph{test artifact}. We retain actions with logged environment return codes, excluding rejected action formats. A shared deterministic pipeline detects commands that write complete files through heredocs, \texttt{echo}/\texttt{printf} redirection, and Python writes whose paths and contents can be recovered without executing the command. It keeps Python filenames starting with \texttt{test} or ending with \texttt{\_test.py} or \texttt{tests.py}, extending common test-discovery conventions~\cite{pytest_test_discovery_2026}. Paths are normalized relative to \texttt{/testbed} and counted once per task. Later whole-file writes update the stored content; RQ2 analyzes the last detected version of each file. Files in temporary directories are distinguished by the action and source line that create the directory.

An execution event is a logged bash action that invokes a recognized Python or test runner with a detected file as an argument. Each action counts once, including actions containing multiple commands, and receives its logged return code. This definition measures execution actions rather than individual test cases.
\subsection{Implementation Details}
\label{sec:implementation-details}

We run the 2,000 revised-prompt trajectories with \texttt{mini-SWE-agent} 1.17.2 on Ubuntu 22.04.5 and evaluate their patches with the official SWE-bench \texttt{sb-cli}. The corresponding public baselines use release 1.15.0 for Gemini, 1.17.1 for Kimi and DeepSeek, and 1.17.2 for GPT-5.2. Revised GPT and Gemini runs use OpenRouter; Kimi and DeepSeek use their provider APIs. The released trajectory metadata records the exact model identifiers, decoding settings, prompts, and run limits for both conditions. The revised runs cost approximately USD~1{,}600 in total.

%% file: images/study_design_overview.tex
\begin{tikzpicture}[
  x=1cm,
  y=1cm,
  card/.style={draw=black!24, fill=white, line width=0.45pt},
  source/.style={draw=black!24, fill=black!2, line width=0.45pt,
    minimum width=4.58cm, minimum height=0.48cm, text width=4.35cm,
    align=center, inner xsep=4pt, inner ysep=3pt,
    font=\footnotesize, text=black!78},
  focus/.style={line width=0.65pt, minimum width=4.58cm,
    minimum height=0.64cm, text width=4.35cm, align=center,
    inner xsep=4pt, inner ysep=4pt, font=\small\bfseries,
    text=studyink},
  output/.style={draw=black!24, fill=white, line width=0.45pt,
    minimum width=4.58cm, minimum height=0.66cm, text width=4.35cm,
    align=center, inner xsep=4pt, inner ysep=4pt,
    font=\footnotesize\bfseries, text=studyink},
  flow/.style={-{Latex[length=1.55mm,width=1.05mm]}, draw=black!45,
    line width=0.55pt}
]
\definecolor{studyink}{HTML}{263746}
\definecolor{rqblue}{HTML}{326B91}
\definecolor{rqteal}{HTML}{2F7D73}
\definecolor{rqorange}{HTML}{B65E3C}

\path[card] (-8.15,-1.72) rectangle (-2.90,1.80);
\fill[rqblue] (-8.15,1.72) rectangle (-2.90,1.80);
\path[card] (-2.63,-1.72) rectangle (2.63,1.80);
\fill[rqteal] (-2.63,1.72) rectangle (2.63,1.80);
\path[card] (2.90,-1.72) rectangle (8.15,1.80);
\fill[rqorange] (2.90,1.72) rectangle (8.15,1.80);

\node[anchor=west, font=\fontsize{13}{13}\selectfont\bfseries, text=rqblue]
  at (-7.86,1.39) {RQ1};
\node[anchor=west, font=\small\bfseries, text=studyink]
  at (-6.87,1.39) {Testing behavior};
\draw[black!17] (-7.86,1.10) -- (-3.19,1.10);

\node[source] (source1) at (-5.525,0.76) {Agent trajectories};
\node[focus, draw=rqblue, fill=rqblue!5] (focus1) at (-5.525,-0.02)
  {Test writing and execution};
\node[output] (output1) at (-5.525,-1.03)
  {Frequency\,$\cdot$\,Timing\,$\cdot$\,Execution};
\draw[flow] (source1.south) -- (focus1.north);
\draw[flow] (focus1.south) -- (output1.north);

\node[anchor=west, font=\fontsize{13}{13}\selectfont\bfseries, text=rqteal]
  at (-2.34,1.39) {RQ2};
\node[anchor=west, font=\small\bfseries, text=studyink]
  at (-1.35,1.39) {Test-file content};
\draw[black!17] (-2.34,1.10) -- (2.34,1.10);

\node[source] (source2) at (0,0.76) {Agent-written test files};
\node[focus, draw=rqteal, fill=rqteal!5] (focus2) at (0,-0.02)
  {AST classification};
\node[output] (output2) at (0,-1.03)
  {Assertions\,$\cdot$\,Value-revealing prints};
\draw[flow] (source2.south) -- (focus2.north);
\draw[flow] (focus2.south) -- (output2.north);

\node[anchor=west, font=\fontsize{13}{13}\selectfont\bfseries, text=rqorange]
  at (3.19,1.39) {RQ3};
\node[anchor=west, font=\small\bfseries, text=studyink]
  at (4.18,1.39) {Test-writing prompts};
\draw[black!17] (3.19,1.10) -- (7.86,1.10);

\node[source] (source3) at (5.525,0.76) {Baseline vs. revised prompts};
\node[focus, draw=rqorange, fill=rqorange!5] (focus3) at (5.525,-0.02)
  {Encourage or discourage\\new test writing};
\node[output] (output3) at (5.525,-1.03)
  {Test status\,$\cdot$\,Resolution\\API calls + tokens};
\draw[flow] (source3.south) -- (focus3.north);
\draw[flow] (focus3.south) -- (output3.north);
\end{tikzpicture}

%% file: sections/RQ1.tex
\section{RQ1: What Testing Behaviors Emerge Under a Lightweight Agent Scaffold?}
\label{sec:rq1}

\paragraph{Motivation}
RQ1 establishes a descriptive baseline of whether agents write tests, when they introduce them, and how intensively they run them when testing is optional.

\paragraph{Experiment Design}
We compare resolved and unresolved trajectories descriptively across test-writing frequency, timing, and execution; these comparisons do not estimate a causal effect of testing on success.
We use two-sided Mann--Whitney U tests for comparisons within each model. For comparisons combining all models (pooled), task-clustered permutation tests keep the six models' outcomes for the same task together. Holm correction is applied within each table.
Pooled means give each eligible task--model run equal weight.

\subsection{RQ1.1 Frequency: How Often Do Agents Write Test Files?}
\label{sec:rq1-test-writing}
\paragraph{Goal and measurements}
We examine whether models write test artifacts under a lightweight scaffold.
For each task, we record (i) whether the agent writes at least one test artifact, and (ii) if so, how many distinct test artifacts it writes.
We report results separately for resolved and unresolved tasks.
For a given model and outcome (resolved or unresolved), let $\mathcal{T}$ be the set of tasks and $A_t$ the number of distinct test files written in task $t$. The subset $\mathcal{T}^{+}=\{t\in\mathcal{T}:A_t>0\}$ contains tasks with at least one test file.
We average the file count over this subset:
\[
    \text{Mean \#test files}=\frac{1}{|\mathcal{T}^{+}|}\sum_{t\in\mathcal{T}^{+}} A_t.
\]

\begin{table*}[t]
\centering
\small
\caption{Per-model test writing rate by task-resolution outcome}
\label{tab:rq1-1-testgen-permodel}
\begin{adjustbox}{max width=\textwidth}
\setlength{\tabcolsep}{2.6pt}%
\renewcommand{\arraystretch}{1.03}%
\begin{tabular}{l c c c c c c c c c}
\toprule
& \multicolumn{3}{c}{Resolved} & \multicolumn{3}{c}{Unresolved} & \multicolumn{3}{c}{All} \\
\cmidrule(lr){2-4}\cmidrule(lr){5-7}\cmidrule(lr){8-10}
Model
& \makecell{\#Tasks}
& \makecell{Tasks w/\\test files}
& \makecell{Mean\\\#test files}
& \makecell{\#Tasks}
& \makecell{Tasks w/\\test files}
& \makecell{Mean\\\#test files}
& \makecell{\#Tasks}
& \makecell{Tasks w/\\test files}
& \makecell{Mean\\\#test files} \\
\midrule
\textit{Claude Opus 4.5} & 372 & 316 (84.9\%) & 3.32 & 128 & 102 (79.7\%) & 4.15 & 500 & 418 (83.6\%) & 3.52 \\
\textit{Gemini 3 Pro} & 371 & 236 (63.6\%) & 2.00 & 129 & 74 (57.4\%) & 2.12 & 500 & 310 (62.0\%) & 2.03 \\
\textit{GPT-5.2} & 359 & 9 (2.5\%) & 1.56 & 141 & 3 (2.1\%) & 1.67 & 500 & 12 (2.4\%) & 1.58 \\
\textit{Kimi K2 Thinking} & 317 & 309 (97.5\%) & 3.47 & 183 & 179 (97.8\%) & 3.83 & 500 & 488 (97.6\%) & 3.60 \\
\textit{MiniMax M2} & 305 & 302 (99.0\%) & 4.83 & 195 & 191 (97.9\%) & 5.77 & 500 & 493 (98.6\%) & 5.20 \\
\textit{DeepSeek v3.2 Reasoner} & 300 & 277 (92.3\%) & 3.47 & 200 & 165 (82.5\%) & 4.00 & 500 & 442 (88.4\%) & 3.67 \\
\bottomrule
\end{tabular}%
\end{adjustbox}

\vspace{4pt}
\scriptsize\emph{Notes.}
``Tasks w/ test files'' reports count and percentage within each outcome split.
``Mean \#test files'' averages file counts over tasks with test files; each distinct file path is counted once per task.
\end{table*}

\paragraph{Results}
Table~\ref{tab:rq1-1-testgen-permodel} reports test-writing frequency by model and outcome.
\textit{GPT-5.2} writes detected test files in only 12/500 tasks.
Within-model differences between resolved and unresolved tasks are smaller than the cross-model variation.
When tests are written, agents on unresolved tasks often write as many or more distinct test artifacts than agents on resolved tasks.

\begin{takeawaybox}[RQ1.1 Test Writing: Key Finding]
Writing new test files more often does not necessarily mean solving more issues. On the same 500 issues, GPT-5.2 creates new test files in 2.4\% of its attempts and solves 71.8\% of the issues. Kimi K2 Thinking creates new test files in 97.6\% of its attempts but solves 63.4\% of the issues.
\end{takeawaybox}

\subsection{RQ1.2 Timing: When Are Tests Written During the Run?}
\label{sec:rq1-timing}
\paragraph{Goal and measurements}
We measure the normalized position of the \textbf{first} and \textbf{last} test-writing steps and their \textbf{span} among tasks that write tests. Because \textit{GPT-5.2} writes tests in only 12 tasks, we omit it from per-model analyses that require test writing.

\begin{align*}
    t_{\text{first}} &= \frac{\min(S_{\text{write}})}{N_{\text{steps}}}, \quad 
    t_{\text{last}}  = \frac{\max(S_{\text{write}})}{N_{\text{steps}}} \\[6pt]
    s_{\text{write}} &= t_{\text{last}} - t_{\text{first}}
    = \frac{\max(S_{\text{write}}) - \min(S_{\text{write}})}{N_{\text{steps}}}
\end{align*}

Here, $S_{\text{write}}$ contains zero-based assistant-step indices of detected file writes, including rewrites, and $N_{\text{steps}}$ counts all assistant messages in the trajectory. The measures lie in $[0,1]$; larger positions indicate later writing. A run with only one detected writing step has zero span. The span measures the separation between writing events, rather than time spent testing.

\begin{table*}[t]
\centering
\small
\caption{Per-model timing of test writing events}
\label{tab:rq1-2-timing-combined}
\begin{adjustbox}{max width=\textwidth}
\setlength{\tabcolsep}{2.5pt}%
\renewcommand{\arraystretch}{1.03}%
\newcommand{\rqonetwopad}{\hphantom{$^{***}$}}
\begin{tabular}{l
                c c c c
                c c c c}
\toprule
& \multicolumn{4}{c}{Resolved} & \multicolumn{4}{c}{Unresolved} \\
\cmidrule(lr){2-5}\cmidrule(lr){6-9}
Model
& \makecell{\#Tasks}
& \makecell{First\\pos.}
& \makecell{Last\\pos.}
& \makecell{Span}
& \makecell{\#Tasks}
& \makecell{First\\pos.}
& \makecell{Last\\pos.}
& \makecell{Span} \\
\midrule
\textit{Claude Opus 4.5} & 316 & 0.34 & 0.75 & 0.41 & 102 & 0.30 & 0.78 & 0.48 \\
\textit{Gemini 3 Pro} & 236 & 0.54 & 0.67 & 0.13 & 74 & 0.55 & 0.69 & 0.14 \\
\textit{Kimi K2 Thinking} & 309 & 0.40 & 0.82 & 0.42 & 179 & 0.40 & 0.82 & 0.42 \\
\textit{MiniMax M2} & 302 & 0.35 & 0.86 & 0.51 & 191 & 0.28$^{**}$ & 0.85$^{*}$ & 0.57 \\
\textit{DeepSeek v3.2 Reasoner} & 277 & 0.43 & 0.80 & 0.37 & 165 & 0.40 & 0.80 & 0.40 \\
\midrule
\textbf{All models} & 1449 & 0.41 & 0.78 & 0.38 & 714 & 0.37$^{*}$ & 0.80$^{*}$ & 0.43$^{*}$ \\
\bottomrule
\end{tabular}%
\end{adjustbox}

\vspace{4pt}
\footnotesize\emph{Notes.}
Means over runs with test files; positions and spans are fractions of each run's total interaction steps. GPT-5.2 is included only in the pooled row.
Pooled tests jointly permute six-model outcome vectors across tasks. Asterisks use Holm-adjusted $p$ values across 15 per-model and three pooled tests: $^{*}p<0.05$, $^{**}p<0.01$, $^{***}p<0.001$.
\end{table*}

\paragraph{Results}
Table~\ref{tab:rq1-2-timing-combined} shows that the mean last writing position lies late in the trajectory (0.67--0.86). \textit{Gemini 3 Pro} starts later and has the shortest span (0.13--0.14), while \textit{MiniMax M2} has the widest span (0.51--0.57).

\begin{takeawaybox}[RQ1.2 Writing Timing: Key Finding]
On unsolved issues, test-file writing extends over a larger portion of the repair process. For attempts with test files, we measure the steps from the first test-file write to the last as a share of all interaction steps. Combining the six models, this share averages 43\% for unsolved issues and 38\% for solved issues.
\end{takeawaybox}

\subsection{RQ1.3 Execution: How Intensively Are Agent-Written Tests Executed, and With What Action Outcomes?}
\label{sec:rq1-exec}
\paragraph{Goal and measurements}
RQ1.3 describes how agents execute tests after they have written them.
We measure (i) the number of test-execution actions, (ii) execution actions per written artifact, and (iii) the fraction of these actions with a non-zero exit code.
An action can fail because of a test, an environment error, or another command in the same action; its return code does not isolate assertion failures.

For each task $t$, let $E_t$ be the number of test-execution actions, $A_t$ the number of agent-written test artifacts, and $Z_t$ the number of these actions with non-zero return codes.
We report three task-level metrics:
\textbf{execution actions per task} ($E_t$);
\textbf{execution actions per test file} ($E_t/A_t$);
and \textbf{nonzero-exit rate} ($Z_t/E_t$), the fraction of executions with a non-zero return code.
We compute each metric per task and then average the resulting values. The first two metrics use all tasks with test artifacts; the nonzero-exit rate uses tasks with $E_t>0$, because the ratio is undefined when no test execution is detected.

\begin{table*}[!t]
\centering
\small
\caption{Task-level execution effort and action-level outcomes of agent-written tests}
\label{tab:rq1-3-task-exec}
\begin{adjustbox}{max width=\textwidth}
\setlength{\tabcolsep}{2.6pt}%
\renewcommand{\arraystretch}{1.05}%
\begin{tabular}{l
                r r r r
                r r r r}
\toprule
& \multicolumn{4}{c}{Resolved} & \multicolumn{4}{c}{Unresolved} \\
\cmidrule(lr){2-5}\cmidrule(lr){6-9}
Model
& \makecell{\#Tasks w/\\test files}
& \makecell{Exec. actions\\per task}
& \makecell{Exec. actions\\per test file}
& \makecell{Nonzero-exit\\rate (\%)}
& \makecell{\#Tasks w/\\test files}
& \makecell{Exec. actions\\per task}
& \makecell{Exec. actions\\per test file}
& \makecell{Nonzero-exit\\rate (\%)} \\
\midrule
\textit{Claude Opus 4.5} & 316 & 4.82 & 1.48 & 12.11 & 102 & 6.07$^{*}$ & 1.58 & 10.59 \\
\textit{Gemini 3 Pro} & 236 & 2.67 & 1.50 & 8.06 & 74 & 2.77 & 1.48 & 7.52 \\
\textit{Kimi K2 Thinking} & 309 & 5.37 & 1.62 & 24.92 & 179 & 6.49 & 1.75 & 21.07 \\
\textit{MiniMax M2} & 302 & 7.14 & 1.54 & 24.03 & 191 & 9.67$^{***}$ & 2.08 & 24.33 \\
\textit{DeepSeek v3.2 Reasoner} & 277 & 3.64 & 1.10 & 28.38 & 165 & 4.64 & 1.36 & 31.37 \\
\midrule
\textbf{All models} & 1449 & 4.82 & 1.44 & 19.75 & 714 & 6.45$^{***}$ & 1.69 & 21.32 \\
\bottomrule
\end{tabular}%
\end{adjustbox}

\vspace{4pt}
\scriptsize
\emph{Notes.}
\#Tasks: tasks with test writing; all other values are averages of the per-task metrics.
Nonzero-exit rates exclude tasks with no detected execution action. The pooled row includes GPT-5.2.
Pooled tests jointly permute six-model outcome vectors across tasks. Asterisks use Holm-adjusted $p$ values across 15 per-model and three pooled tests: $^{*}p<0.05$, $^{**}p<0.01$, $^{***}p<0.001$.
\end{table*}

\paragraph{Results}
Table~\ref{tab:rq1-3-task-exec} summarizes execution effort and action-level outcomes for tasks that write tests.
In the pooled data, unresolved tasks have more execution actions per test file than resolved tasks (1.69 vs.\ 1.44), although this difference is not significant after Holm correction.
At the model level, the difference in execution actions per task remains significant for \textit{Claude Opus 4.5} and \textit{MiniMax M2}; no model-level difference in actions per test file or nonzero-exit rate remains significant.

The nonzero-exit rate is slightly higher for unresolved tasks (21.32\% vs.\ 19.75\%), but this pooled difference is not statistically significant. \textit{Gemini 3 Pro} has the fewest execution actions (about 2.7--2.8 per task), while \textit{MiniMax M2} has the most, especially for unresolved tasks (9.67 per task; 2.08 per test file). Nonzero-exit rates are lower for \textit{Claude Opus 4.5} and \textit{Gemini 3 Pro} (about 8--12\%) than for the other three models (about 21--31\%).

\begin{takeawaybox}[RQ1.3 Test Execution: Key Finding]
Agents run newly written tests more often on issues they fail to solve. Among attempts that create test files, we count tool calls that run these files. Combining the six models, the average is 6.45 such calls per unsolved issue and 4.82 per solved issue.
\end{takeawaybox}

%% file: sections/RQ2.tex
\section{RQ2: What Assertion and Value-Revealing Print Patterns Appear in Agent-Written Tests?}
\label{sec:rq2}

\paragraph{Motivation}
Agent-written test files can encode both explicit checks and statements that expose runtime information.
RQ1 treats tests as \emph{events} in the trajectory, measuring whether agents write them, when they appear, and how often they are run.
RQ2 shifts to the \emph{content} of those files and characterizes two common statement types:
\textbf{assertions}, which specify conditions to check, and \textbf{value-revealing prints}, which expose runtime values or execution information.

\paragraph{Experiment Design}
RQ2 examines tasks in which the agent writes at least one test file and summarizes three aspects of file content:
\begin{itemize}[leftmargin=*]
    \item \textbf{Statement counts} (RQ2.1): assertion vs.\ value-revealing print counts in agent-written tests.
    \item \textbf{Assertion types} (RQ2.2): what types of assertions appear in agent-written tests, using a four-type categorization.
    \item \textbf{Print types} (RQ2.3): what runtime information value-revealing prints expose, using a three-type categorization.
\end{itemize}
RQ2.1 stays at the \emph{task level} and measures aggregate assertion and print counts. RQ2.2 and RQ2.3 then move to the \emph{statement level}, categorizing assertion and value-revealing print statements separately.

\subsection{RQ2.1 Task-level statement counts: What statements do agent-written tests contain?}
\label{sec:rq2-1-signal-amount}

\paragraph{Goal and measurements}
For tasks with at least one agent-written test file, we count the assertions and value-revealing prints in those files.
We distinguish two statement types: (i) \textbf{assertions} ($A$), i.e., native \texttt{assert} statements and recognized framework assertion calls that specify explicit checks, and
(ii) \textbf{value-revealing prints} ($P$), i.e., \texttt{print} statements with at least one argument that depends on a variable, attribute, indexed value, runtime expression, or function call. We exclude prints composed only of literals, including fixed messages and expected-value text (e.g., \texttt{print("Expected: 3")} and \texttt{print("=" * 60)}), while retaining statements such as \texttt{print(obj.attr)}.
Because GPT-5.2 produces test artifacts in only 12 of 500 baseline tasks, we exclude it from the cross-model content summaries in RQ2. The same AST rule is used for the task-level counts in RQ2.1 and the print categories in RQ2.3. Of the 8,041 artifacts from the five models analyzed in RQ2, 8,028 parse successfully; the remaining 13 contribute no assertion or print statements to these analyses.
For each task $t$ with test artifacts $\mathcal{A}_t$, and for statement type $S \in \{A,P\}$, let $n^S_{t,a}$ be the number of statements of type $S$
in artifact $a \in \mathcal{A}_t$. We define the \textbf{task-level statement totals}:
\begingroup\small
\[
N^{S}_t=\sum_{a\in\mathcal{A}_t} n^{S}_{t,a},
\qquad
N^{\mathrm{total}}_t = N^{A}_t + N^{P}_t.
\]
\endgroup
For each model, we average $N^A_t$ (assertion count), $N^P_t$ (value-revealing print count), and $N^{\mathrm{total}}_t$ (overall statement count) across tasks, giving each task equal weight and reporting resolved and unresolved tasks separately.

\begin{table*}[t]
\centering
\small
\caption{Task-level assertion and value-revealing print counts in agent-written tests}
\label{tab:rq2-1-signal-amount}

\begin{adjustbox}{max width=\textwidth}
\begin{tabular}{l c c c c c c c c}
\toprule
& \multicolumn{4}{c}{Resolved} & \multicolumn{4}{c}{Unresolved} \\
\cmidrule(lr){2-5}\cmidrule(lr){6-9}
Model
& \makecell{\#Tasks\\w/ tests}
& \makecell{Assertions\\per task\\\footnotesize($\bar{N}^{A}$)}
& \makecell{Prints\\per task\\\footnotesize($\bar{N}^{P}$)}
& \makecell{Assertions + prints\\per task\\\footnotesize($\bar{N}^{\mathrm{total}}$)}
& \makecell{\#Tasks\\w/ tests}
& \makecell{Assertions\\per task\\\footnotesize($\bar{N}^{A}$)}
& \makecell{Prints\\per task\\\footnotesize($\bar{N}^{P}$)}
& \makecell{Assertions + prints\\per task\\\footnotesize($\bar{N}^{\mathrm{total}}$)} \\
\midrule
\textit{Claude Opus 4.5} & 316 & 5.12 & 19.96 & 25.08 & 102 & 5.30 & 21.14 & 26.44 \\
\textit{Gemini 3 Pro} & 236 & 1.47 & 4.30 & 5.77 & 74 & 1.59 & 4.97 & 6.57 \\
\textit{Kimi K2 Thinking} & 309 & 2.87 & 18.96 & 21.83 & 179 & 3.58 & 21.78 & 25.36 \\
\textit{MiniMax M2} & 302 & 7.36 & 30.73 & 38.10 & 191 & 4.66 & 39.81 & 44.48 \\
\textit{DeepSeek v3.2 Reasoner} & 277 & 3.57 & 15.86 & 19.43 & 165 & 3.27 & 20.10 & 23.37 \\
\bottomrule
\end{tabular}
\end{adjustbox}

\vspace{2pt}
\parbox{\textwidth}{\footnotesize
\textbf{Note.}
Means over tasks with test files, giving each task equal weight.
Assertions count native \texttt{assert} statements and recognized framework assertion calls; prints count \emph{value-revealing} \texttt{print}s.
$\bar{N}^{\mathrm{total}}=\bar{N}^{A}+\bar{N}^{P}$.
}
\end{table*}

\paragraph{Results}
Table~\ref{tab:rq2-1-signal-amount} shows marked variation in statement volume, from the lower totals of \textit{Gemini 3 Pro} to the higher totals of \textit{MiniMax M2}.

\begin{takeawaybox}[RQ2.1 Statement Counts: Key Finding]
Agent-written test files contain more statements that show runtime information than statements that check conditions automatically. For example, \texttt{print(result)} shows a value; \texttt{assert result == expected} checks it. After totaling these statements across the test files for each issue, the average print count exceeds the average assertion count for all five models, on both solved and unsolved issues.
\end{takeawaybox}

\subsection{RQ2.2 Assertion categorization: What kinds of verification do assertions encode?}
\label{sec:rq2-2-assertion-categorization}

\paragraph{Goal and measurements}
RQ2.1 counts assertion statements in agent-written tests, but counts alone do not tell us \emph{what} those assertions check.
Assertions can enforce basic preconditions (e.g., non-\texttt{None} or type checks) or compare results with expected values or structures.
Models may therefore differ not only in how often they assert, but also in what forms of checks they write. RQ2.2 provides a descriptive breakdown of \texttt{assert} statements into four assertion categories:

\begin{itemize}[leftmargin=*]
  \item \textbf{C1 Sanity checks.}
  The assertion checks truthiness, identity, existence, or type.
  \emph{Example:} \texttt{assert x is not None}.

  \item \textbf{C2 Property checks.}
  The assertion checks a property of a value or object (e.g., membership or validity) without fixing an exact output.
  \emph{Example:} \texttt{assert key in mapping}.

  \item \textbf{C3 Relational checks.}
  The assertion enforces a range, bound, inequality, or approximate relationship between values. \emph{Example:} \texttt{assert 0 <= score <= 1}.
  This category also includes checks that expect a specific exception, because they constrain the allowed behavior to ``must fail with an exception of type $E$'' rather than matching a single concrete output.

  \item \textbf{C4 Exact checks.}
  The assertion checks equality with an expected value or structure, whether represented by a literal or a variable.
  \emph{Example:} \texttt{assert output == expected\_output}.
\end{itemize}

To identify and categorize assertions, we implement a rule-based classifier over Python ASTs and map each assertion to exactly one category.
The classifier covers both native \texttt{assert} statements (e.g., \texttt{assert a == b}) and framework-provided assertion calls (e.g., \texttt{self.assertEqual(a, b)} in \texttt{unittest}). Concretely, for each test artifact, we parse the code into an AST and extract:
(i) native \texttt{assert <expr>} statements, and
(ii) calls to framework assertion APIs. Some \texttt{assert} statements contain multiple checks joined by Boolean operators (e.g., \texttt{assert a > 0 and b == 1}). For these compound statements, we recursively split the expression into atomic checks and assign the highest-numbered category present. This keeps one observation per source statement; the category numbers define the tie-breaking rule, not an empirical ranking of assertion quality. Thus, the example receives C4 because \texttt{a > 0} is C3 and \texttt{b == 1} is C4.

We check three alternative ways to classify compound assertions: choosing the lowest-numbered category, using the first check's category, or counting each check separately. Compound Boolean expressions occur in 161 of 8,139 native \texttt{assert} statements (2.0\%; 1.8\% of all 8,796 assertions). Across the three alternatives, C4 remains the largest category (56.96--57.09\%) and C3 the smallest (5.09--5.17\%).

\begin{table*}[t]
\centering
\small
\caption{Assertion category distribution by model.
Counts and percentages are computed over all detected assertion statements and calls written by each model.}
\label{tab:rq2-2-assertion-categories}

\begin{adjustbox}{max width=\textwidth}
\begin{tabular}{l r r r r r r r r r}
\toprule
Model & \#Assertions
& \multicolumn{2}{c}{C1 Sanity}
& \multicolumn{2}{c}{C2 Property}
& \multicolumn{2}{c}{C3 Relational}
& \multicolumn{2}{c}{C4 Exact} \\
\cmidrule(lr){3-4}
\cmidrule(lr){5-6}
\cmidrule(lr){7-8}
\cmidrule(lr){9-10}
& & \# & \% & \# & \% & \# & \% & \# & \% \\
\midrule
\textit{Claude Opus 4.5} & 2159 & 353 & 16.35\% & 430 & 19.92\% & 115 & 5.33\% & 1261 & 58.41\% \\
\textit{Gemini 3 Pro} & 466 & 78 & 16.74\% & 105 & 22.53\% & 43 & 9.23\% & 240 & 51.50\% \\
\textit{Kimi K2 Thinking} & 1527 & 227 & 14.87\% & 338 & 22.13\% & 54 & 3.54\% & 908 & 59.46\% \\
\textit{MiniMax M2} & 3115 & 618 & 19.84\% & 595 & 19.10\% & 176 & 5.65\% & 1726 & 55.41\% \\
\textit{DeepSeek v3.2 Reasoner} & 1529 & 289 & 18.90\% & 261 & 17.07\% & 66 & 4.32\% & 913 & 59.71\% \\
\bottomrule
\end{tabular}
\end{adjustbox}

\vspace{2pt}
\parbox{\textwidth}{\footnotesize
\textbf{Note.}
C1--C4 denote four assertion categories defined by the form of the check (sanity, property, relational/approximate, and exact-output).
Percentages are computed within each model, relative to the model's total assertion count.
}
\end{table*}

\paragraph{Results}
Table~\ref{tab:rq2-2-assertion-categories} reports assertion categories by model. \textbf{C2 Property} and \textbf{C1 Sanity} account for 17.07--22.53\% and 14.87--19.84\% of assertions, respectively.

\begin{takeawaybox}[RQ2.2 Assertion Types: Key Finding]
In agent-written test files, most assertions require an exact expected value, as in \texttt{assert output == expected} (51.5--59.7\% in each of the five models). Assertions that accept a range of values, accept a value close to the expected one, or check for an expected exception are much less common (3.5--9.2\% combined).
\end{takeawaybox}

\subsection{RQ2.3 Print categorization: What information do value-revealing prints expose?}
\label{sec:rq2-3-print-categorization}

\paragraph{Goal and measurements}
Since value-revealing prints substantially outnumber assertions, RQ2.3 categorizes the information exposed by these statements as concrete values or content, structural summaries, or exception and execution-status information:

\begin{itemize}[leftmargin=*]
  \item \textbf{P1 Value / content inspection.}
  The print exposes a concrete runtime value or content, such as a returned output, an intermediate result, an object field, or part of a generated string. This category captures \emph{what the program produced}, rather than a structural summary or an error/status signal;
  \emph{e.g.,} \texttt{print(add(1, 2))}.

  \item \textbf{P2 Structural summary inspection.}
  The print reports an aggregate structural summary, such as length, size, shape, number of items, or emptiness. This category is reserved for prints that summarize \emph{how much} data is present or \emph{what structure} it has, rather than showing the content itself;
  \emph{e.g.,} \texttt{print(len(items))}.

  \item \textbf{P3 Exception / execution-status signals.}
  The print surfaces an exception, an error message, or a coarse execution-status indicator, such as a success/failure flag. This category is used when the print primarily indicates \emph{what happened during execution}, rather than the program's computed content;
  \emph{e.g.,} \texttt{except Exception as exc: print(exc)}.
\end{itemize}

We implement a rule-based classifier over Python ASTs. Using the same inclusion rule as RQ2.1, we extract value-revealing \texttt{print(...)} calls and apply the categories in a fixed order: P3 for prints in exception handlers or with exception/status cues, P2 for size-, shape-, or count-oriented expressions, and P1 for the remaining included prints. Table~\ref{tab:rq2-3-print-categories} reports the resulting category distribution.

\begin{table*}[t]
\centering
\small
\caption{Value-revealing print category distribution by model.}
\label{tab:rq2-3-print-categories}

\begin{adjustbox}{max width=\textwidth}
\begin{tabular}{l r r r r r r r}
\toprule
Model & \#Prints
& \multicolumn{2}{c}{P1 Value/Content}
& \multicolumn{2}{c}{P2 Structural}
& \multicolumn{2}{c}{P3 Exception/Status} \\
\cmidrule(lr){3-4}
\cmidrule(lr){5-6}
\cmidrule(lr){7-8}
& & \# & \% & \# & \% & \# & \% \\
\midrule
\textit{Claude Opus 4.5} & 8462 & 6595 & 77.94\% & 285 & 3.37\% & 1582 & 18.70\% \\
\textit{Gemini 3 Pro} & 1382 & 969 & 70.12\% & 72 & 5.21\% & 341 & 24.67\% \\
\textit{Kimi K2 Thinking} & 9758 & 7271 & 74.51\% & 644 & 6.60\% & 1843 & 18.89\% \\
\textit{MiniMax M2} & 16885 & 12065 & 71.45\% & 958 & 5.67\% & 3862 & 22.87\% \\
\textit{DeepSeek v3.2 Reasoner} & 7709 & 5906 & 76.61\% & 288 & 3.74\% & 1515 & 19.65\% \\
\bottomrule
\end{tabular}
\end{adjustbox}

\end{table*}

\paragraph{Results}
Table~\ref{tab:rq2-3-print-categories} categorizes all 44,196 value-revealing prints counted in RQ2.1. \textbf{P3 Exception / execution-status signals} accounts for 18.70--24.67\%, while \textbf{P2 Structural summary inspection} accounts for 3.37--6.60\%.

\begin{takeawaybox}[RQ2.3 Print Types: Key Finding]
When agents print runtime information in test files, they mainly show function results, intermediate values, or object fields. These account for 70.1--77.9\% of such prints in each of the five models. Printing an error or reporting whether an operation succeeded is less common; printing summaries such as length or shape is least common.
\end{takeawaybox}

%% file: sections/RQ3.tex
\section{RQ3: How Do Outcomes and Resource Use Differ Across Test-Writing Prompt Conditions?}
\label{sec:rq3}

\paragraph{Motivation}
RQ1 shows that models with similar issue-resolution rates can differ sharply in how often they write test files. RQ2 shows that these files contain more value-revealing prints than assertions. RQ3 examines outcomes and resource use under prompts that encourage or discourage new test files.

\paragraph{Experiment Design}
RQ3 compares task-resolution outcomes (RQ3.1) and resource use (RQ3.2) between baseline and revised prompt conditions.

\paragraph{Model selection}
We select models by their baseline test-writing rates (Table~\ref{tab:rq1-1-testgen-permodel}). The encourage-test comparison uses \textit{GPT-5.2} (2.4\%) and \textit{Gemini 3 Pro} (62.0\%), whose baselines leave room for more test creation. For the discourage-test comparison, the experiment budget supported two models; we selected \textit{Kimi K2 Thinking} (97.6\%) and \textit{DeepSeek v3.2 Reasoner} (88.4\%) from the high test-writing group.

We use the mini-SWE-agent prompt to create two variants:
\begin{itemize}[leftmargin=*]
    \item \textbf{Encourage writing tests:} for \textit{GPT-5.2} and \textit{Gemini 3 Pro}, we append prompt instructions to write at least \emph{one} runnable new test file (a file whose name starts with \texttt{test\_} or ends with \texttt{\_test.py}), separate from the repository's existing tests.
    \item \textbf{Discourage writing tests:} for \textit{Kimi K2 Thinking} and \textit{DeepSeek v3.2 Reasoner}, we remove the default testing-related prompt cue from mini-SWE-agent and append prompt instructions not to write any new test files or scripts.
\end{itemize}
Reference prompt templates are included in \textit{experiment\allowbreak\_prompts}; each trajectory records its exact prompt. Gemini's encouragement prompt repeats the instruction in action feedback, whereas GPT-5.2's uses system and instance instructions. We compare each revised condition with its public baseline on the same tasks.

\subsection{RQ3.1 How do task outcomes differ across test-writing prompt conditions?}
\label{sec:rq3-1-outcomes}

\paragraph{Goal and measurements} For each model, we compare the same 500 tasks under the baseline and revised prompts. We record whether each run creates a test artifact (\emph{No test} or \emph{Has test}) and resolves the issue (\emph{Fail} or \emph{Success}), then cross-tabulate their baseline-to-revised-prompt transitions. Runs in either test-writing group can also execute existing repository tests or inline checks. We use exact McNemar tests with Holm correction across the four models and 95\% paired bootstrap confidence intervals to compare resolution rates.

\begin{table*}[t]
\centering
\small
\caption{Task-level test-status and outcome transitions under the revised prompts}
\label{tab:rq3-outcome-transition}
\begin{adjustbox}{max width=\textwidth}
\begin{tabular}{lccccc}
\toprule
\multicolumn{6}{l}{\textbf{Encourage writing tests}} \\
\midrule
Outcome transition
& No test $\rightarrow$ No test
& \cellcolor{blue!12}\textbf{No test $\rightarrow$ Has test}
& Has test $\rightarrow$ No test
& Has test $\rightarrow$ Has test
& Total \\
\midrule

\multicolumn{2}{l}{\textbf{Model: GPT-5.2} ($p_{\mathrm{H}}=1.000$)}
& \cellcolor{blue!7}$\Delta$ \textbf{466 (93.2\%)} & & & \\
Fail $\rightarrow$ Success & 1 & \cellcolor{blue!12}24 & 0 & 2 & 27 \\
Success $\rightarrow$ Fail & 3 & \cellcolor{blue!12}24 & 0 & 0 & 27 \\
Stable Success & 8 & \cellcolor{blue!12}315 & 0 & 9 & 332 \\
Stable Fail & 10 & \cellcolor{blue!12}103 & 0 & 1 & 114 \\
\midrule
\textbf{Net change in \#Success}
& \textbf{$-$2}
& \cellcolor{blue!8}\textbf{0}
& \textbf{0}
& \textbf{2}
& \textbf{0} \\
\midrule\midrule

\multicolumn{2}{l}{\textbf{Model: Gemini 3 Pro} ($p_{\mathrm{H}}=1.000$)}
& \cellcolor{blue!7}$\Delta$ \textbf{184 (36.8\%)} & & & \\
Fail $\rightarrow$ Success & 0 & \cellcolor{blue!12}8 & 0 & 9 & 17 \\
Success $\rightarrow$ Fail & 1 & \cellcolor{blue!12}10 & 1 & 10 & 22 \\
Stable Success & 2 & \cellcolor{blue!12}122 & 4 & 221 & 349 \\
Stable Fail & 3 & \cellcolor{blue!12}44 & 0 & 65 & 112 \\
\midrule
\textbf{Net change in \#Success}
& \textbf{$-$1}
& \cellcolor{blue!8}\textbf{$-$2}
& \textbf{$-$1}
& \textbf{$-$1}
& \textbf{$-$5} \\
\midrule\midrule

\multicolumn{6}{l}{\textbf{Discourage writing tests}} \\
\midrule
Outcome transition
& No test $\rightarrow$ No test
& No test $\rightarrow$ Has test
& \cellcolor{orange!16}\textbf{Has test $\rightarrow$ No test}
& Has test $\rightarrow$ Has test
& Total \\
\midrule

\multicolumn{3}{l}{\textbf{Model: Kimi K2 Thinking} ($p_{\mathrm{H}}=0.911$)}
& \cellcolor{orange!9}$\Delta$ \textbf{340 (68.0\%)} & & \\
Fail $\rightarrow$ Success & 1 & 0 & \cellcolor{orange!16}31 & 11 & 43 \\
Success $\rightarrow$ Fail & 1 & 0 & \cellcolor{orange!16}42 & 13 & 56 \\
Stable Success & 5 & 2 & \cellcolor{orange!16}188 & 66 & 261 \\
Stable Fail & 2 & 1 & \cellcolor{orange!16}79 & 58 & 140 \\
\midrule
\textbf{Net change in \#Success}
& \textbf{0}
& \textbf{0}
& \cellcolor{orange!10}\textbf{$-$11}
& \textbf{$-$2}
& \textbf{$-$13} \\
\midrule

\multicolumn{3}{l}{\textbf{Model: DeepSeek v3.2 Reasoner} ($p_{\mathrm{H}}=1.000$)}
& \cellcolor{orange!9}$\Delta$ \textbf{373 (74.6\%)} & & \\
Fail $\rightarrow$ Success & 11 & 2 & \cellcolor{orange!16}28 & 7 & 48 \\
Success $\rightarrow$ Fail & 3 & 0 & \cellcolor{orange!16}49 & 5 & 57 \\
Stable Success & 19 & 1 & \cellcolor{orange!16}187 & 36 & 243 \\
Stable Fail & 21 & 1 & \cellcolor{orange!16}109 & 21 & 152 \\
\midrule
\textbf{Net change in \#Success}
& \textbf{8}
& \textbf{2}
& \cellcolor{orange!10}\textbf{$-$21}
& \textbf{2}
& \textbf{$-$9} \\
\bottomrule
\end{tabular}
\end{adjustbox}
\vspace{2pt}
\parbox{\textwidth}{\footnotesize
\textbf{Note.}
\emph{Test status} is defined by whether the run writes at least one test artifact (``Has test'') or writes none (``No test'').
Columns show the baseline$\rightarrow$revised-prompt \emph{test-status transition}; rows show the baseline$\rightarrow$revised-prompt \emph{outcome transition} (Fail/Success).
The highlighted column indicates the \emph{intended} test-status change (blue: No test$\rightarrow$Has test under \textbf{Encourage}; orange: Has test$\rightarrow$No test under \textbf{Discourage});
$\Delta$ reports the number (and percentage) of tasks in that intended-change column. \textbf{Net change(\#Success)} is computed per column as (\#Fail$\rightarrow$Success) $-$ (\#Success$\rightarrow$Fail).
The model-level $p_{\mathrm{H}}$ value reports the exact McNemar test after Holm correction across the four model comparisons.
}
\end{table*}

\begin{figure}[t]
  \centering
  \includegraphics[width=\columnwidth]{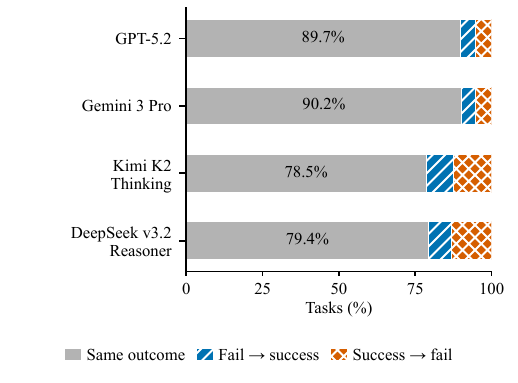}
  \caption{Most tasks with the intended test-status change retain their resolution outcome (78.5\%--90.2\%). The first two bars use the encourage-test prompt; the last two use the discourage-test prompt.}
  \Description{Single-column stacked bars showing that most tasks keep the same final outcome under the revised prompt, with smaller shares changing from fail to success or success to fail. The first two rows show encourage-test conditions and the last two show discourage-test conditions.}
  \label{fig:outcome-transition-stacked-bar}
\end{figure}

\paragraph{Results}
As shown in Table~\ref{tab:rq3-outcome-transition}, the \emph{encourage test writing} condition changes test status for \textit{GPT-5.2} and \textit{Gemini 3 Pro}:
93.2\% and 36.8\% of tasks transition from \emph{No test} to \emph{Has test}, respectively.
Conversely, the \emph{discourage test writing} condition reduces new-test creation for \textit{Kimi K2 Thinking} and \textit{DeepSeek v3.2 Reasoner},
moving 68.0\% and 74.6\% of tasks from \emph{Has test} to \emph{No test}.

Among the 1{,}363 task--model pairs with the intended test-status change across the four comparisons, 1{,}147 (84.2\%) retain the same final outcome (Figure~\ref{fig:outcome-transition-stacked-bar}).
At the model level, resolved-task totals change by 0 for \textit{GPT-5.2}, $-$5 for \textit{Gemini 3 Pro}, $-$13 for \textit{Kimi K2 Thinking}, and $-$9 for \textit{DeepSeek v3.2 Reasoner}.
The corresponding 95\% confidence intervals for the resolution-rate differences, in percentage points, are [$-$2.8, 2.8], [$-$3.4, 1.4], [$-$6.4, 1.2], and [$-$5.8, 2.2].
The exact McNemar tests find no statistically significant baseline-vs.-revised-prompt outcome difference for any model (all Holm-adjusted $p>0.05$).

\begin{takeawaybox}[RQ3.1 Resolution Outcomes: Key Finding]
The same success rate can hide changes in which issues an agent solves. GPT-5.2 solves 359 of 500 issues with the original prompt and 359 with a prompt asking it to write new test files. With the new prompt, 27 issues that were previously unsolved are now solved, but 27 previously solved issues are no longer solved.
\end{takeawaybox}

\subsection{RQ3.2 How do API calls and token usage differ across prompt conditions?}
\label{sec:rq3-2-efficiency}

\begin{table*}[t]
\centering
\small
\caption{Resolution outcomes and resource use under baseline and revised prompt conditions}
\label{tab:rq3-efficiency-tokens}
\begin{adjustbox}{max width=\textwidth}
\setlength{\tabcolsep}{3.5pt}
\begin{tabular}{>{\raggedright\arraybackslash}p{2.2cm}lrrrr}
\toprule
Model & Condition & Tasks resolved & Avg API Calls & Avg Input Tokens & Avg Output Tokens \\
\midrule
\textit{GPT-5.2} &
Baseline & 359 (71.8\%) & 19.76 & 242{,}855 & 24{,}550 \\
&
{Encourage tests} & 359 (71.8\%) & 20.84 & 264{,}762 & 29{,}415 \\
&
Change & \phantom{$-$}+0 (+0.0 pp) & \phantom{$-$}+1.08 (+5.5\%)$^{**}$ & \phantom{$-$}+21{,}907 (+9.0\%)$^{***}$ & \phantom{$-$}+4{,}866 (+19.8\%)$^{***}$ \\
\midrule
\textit{Gemini 3 Pro} &
Baseline & 371 (74.2\%) & 40.33 & 666{,}096 & 11{,}114 \\
&
{Encourage tests} & 366 (73.2\%) & 39.21 & 641{,}307 & 10{,}943 \\
&
Change & $-$5 ($-$1.0 pp) & $-$1.11 ($-$2.8\%) & $-$24{,}789 ($-$3.7\%) & $-$171 ($-$1.5\%) \\
\midrule
\textit{Kimi K2 Thinking} &
Baseline & 317 (63.4\%) & 46.82 & 668{,}449 & 14{,}895 \\
&
{Discourage tests} & 304 (60.8\%) & 30.25 & 340{,}689 & 8{,}468 \\
&
Change & $-$13 ($-$2.6 pp) & $-$16.57 ($-$35.4\%)$^{***}$ & $-$327{,}760 ($-$49.0\%)$^{***}$ & $-$6{,}427 ($-$43.1\%)$^{***}$ \\
\midrule
\textit{DeepSeek v3.2 Reasoner} &
Baseline & 300 (60.0\%) & 46.40 & 637{,}297 & 52{,}120 \\
&
 {Discourage tests} & 291 (58.2\%) & 35.06 & 427{,}780 & 44{,}823 \\
&
Change  & $-$9 ($-$1.8 pp) & $-$11.35 ($-$24.5\%)$^{***}$ & $-$209{,}518 ($-$32.9\%)$^{***}$ & $-$7{,}297 ($-$14.0\%)$^{***}$ \\
\bottomrule
\end{tabular}
\end{adjustbox}

\vspace{2pt}
\parbox{\textwidth}{\footnotesize
\textbf{Note.}
API calls and token counts are per-task means over all 500 tasks for each model and condition. Changes are computed as (Condition $-$ Baseline) before rounding.
For \textbf{Tasks resolved}, parenthesized changes are percentage-point differences; resource-use changes are relative percentages.
\textbf{Encourage tests} is applied to \textit{GPT-5.2} and \textit{Gemini 3 Pro};
\textbf{Discourage tests} is applied to \textit{Kimi K2 Thinking} and \textit{DeepSeek v3.2 Reasoner}.
$^{**}p<0.01$, $^{***}p<0.001$ using Holm-adjusted paired Wilcoxon signed-rank tests across the 12 model--metric comparisons; paired bootstrap CIs are summarized in the text.
}
\end{table*}

\paragraph{Goal and measurements}
We analyze the following three primary metrics based on the trajectories generated in RQ3.1:
(i) average \textit{API calls} per task,
(ii) average \textit{input tokens} per task, and
(iii) average \textit{output tokens} per task.
To distinguish call frequency from call size, we also compute input and output tokens per API call within each task. We average these task-level ratios with equal weight; dividing the overall token mean by the overall call mean would instead weight tasks by their call counts. All five measures are compared on the same 500 paired tasks using 95\% paired bootstrap confidence intervals; the three primary metrics additionally use two-sided Wilcoxon signed-rank tests with Holm correction across the 12 model--metric comparisons.

\paragraph{Results}
Table~\ref{tab:rq3-efficiency-tokens} compares resource use and resolution outcomes across prompt conditions.
Under \emph{encourage test writing}, \textit{GPT-5.2} records higher resource use (+5.5\% API calls; +19.8\% output tokens) with an unchanged resolution total. The paired analyses support this increase across all three metrics: API calls rise by +1.08 per task (95\% CI [0.40, 1.76], Holm-adjusted Wilcoxon $p=0.002$), input tokens by +21{,}907 ([3{,}812, 39{,}918], $p<0.001$), and output tokens by +4{,}866 ([3{,}036, 6{,}701], $p<0.001$). In contrast, \textit{Gemini 3 Pro} changes little overall, and none of its paired differences is statistically significant across API calls, input tokens, or output tokens (all 95\% CIs cross zero).
Under \emph{discourage test writing}, \textit{Kimi K2 Thinking} and \textit{DeepSeek v3.2 Reasoner} record lower resource use: input tokens decrease by 49.0\% and 32.9\%, respectively, and \textit{Kimi K2 Thinking} also uses 35.4\% fewer API calls.
For both models, all three mean differences are negative, all 95\% bootstrap CIs exclude zero, and all Holm-adjusted Wilcoxon tests give $p<0.001$.

We also examine whether token-use differences involve the number of API calls or the tokens used per call. Average input/output tokens per call change by +6.7\%/+21.3\% for \textit{GPT-5.2}, +4.4\%/+0.7\% for \textit{Gemini 3 Pro}, $-$27.5\%/$-$8.6\% for \textit{Kimi K2 Thinking}, and $-$14.8\%/+11.6\% for \textit{DeepSeek v3.2 Reasoner}; paired bootstrap CIs exclude zero except for \textit{Gemini 3 Pro}'s output-token difference. Thus, \textit{Kimi K2 Thinking}'s total-token reductions combine fewer calls with lower per-call token volume, while \textit{DeepSeek v3.2 Reasoner}'s total output-token reduction accompanies fewer calls but higher output tokens per call.
For the two discourage-test comparisons, resource reductions accompany resolution-rate differences of $-$2.6 and $-$1.8 percentage points. The resolution-rate intervals still include potentially meaningful decreases, so these comparisons do not establish equivalent effectiveness.

\begin{takeawaybox}[RQ3.2 Resource Use: Key Finding]
Similar resolution totals can come with substantially different token use. Each model is compared with its original prompt on the same 500 issues. Asked to write new test files, GPT-5.2 uses 9.0\% more input tokens per issue, but its solved total stays at 359. Asked not to create new test files, Kimi K2 Thinking uses 49.0\% fewer input tokens per issue and solves 13 fewer issues; DeepSeek v3.2 Reasoner uses 32.9\% fewer input tokens per issue and solves 9 fewer issues.
\end{takeawaybox}

%% file: sections/discussion.tex
\section{Discussion and Future Work}
\label{sec:discussion}

\subsection{Implications}
Identical resolution totals can hide changes in which issues an agent solves. GPT-5.2 resolves 359 issues under both prompts, but 27 change from failure to success and another 27 change in the opposite direction (Table~\ref{tab:rq3-outcome-transition}). The encouragement condition also uses more tokens. Comparing individual task outcomes and resource use therefore reveals differences that the overall resolution rate misses.

Test-file creation records a workflow choice; resolution measures whether the final patch passes the benchmark's tests. Their different rankings across models make both measurements useful when comparing agents. Reporting whether instructions encourage, discourage, or leave new-test creation to the model also helps explain the setting in which these choices occur.

Agent-written test files combine automatic checks with code for inspecting program behavior. Across the five models in RQ2, value-revealing prints outnumber assertions, and most display concrete values such as returned results or object fields. Assertions encode conditions to check; prints expose information that an agent can inspect. Evaluations should distinguish these two forms of test content.

\subsection{Future work}
A next question is which test outputs lead agents to change their diagnosis or next edit. Trajectory-level analyses can trace these links for new tests, repository tests, and inline commands. Test-quality analyses can also account for transient repository states, where intermediate code and tests may later be overwritten, and examine whether agent-written tests remain useful with the final patch~\cite{yu2023llm,ryan2024code,shin2024domain,bhatia2024unit,yang2024evaluation,MolinelliDGMLEP2025,harman2025mutation}.

\subsection{Threats to Validity}
\label{sec:threats}

\noindent\textbf{\textit{Internal validity.}}
Each prompt condition is observed once per task. Pairing controls task identity, but scaffold releases and API routes differ between some public baselines and revised runs. Decoding and environment variation may also affect outcomes. We therefore interpret the comparisons as observed differences between conditions, without attributing them solely to the prompt edits.

\noindent\textbf{\textit{External validity.}}
Our results cover six LLMs on Python issues from SWE-bench Verified under one lightweight scaffold. Testing conventions, repository languages, and scaffolds with dedicated validation tools may produce different behavior.

\noindent\textbf{\textit{Data construction validity.}}
Our measures cover detected whole-file writes and execution actions; inline-only checks, unmodified repository tests, and indirect file operations fall outside this scope. Local edits may change a file after its last recovered snapshot. Execution return codes describe entire bash actions. AST categories characterize syntax, which may not fully capture semantic intent; static counts do not establish that statements execute or guide later edits. The extraction rules and regression checks are released with the data.

%% file: sections/related_work.tex
\section{Related Work}
\label{sec:related}

\paragraph{LLM-Generated Tests}
Research on LLM-based software testing commonly treats the test artifact as the primary output. Empirical studies evaluate generated unit tests and test suites, while assertion-generation work examines the checks of expected behavior within those tests~\cite{schafer2023empirical,yuan2024evaluating,zhang2025exploring}. Broader evaluations and surveys organize this literature across software-testing tasks and requirements-based test generation~\cite{li2025evaluating,yang2025requirements}. Benchmarks make these targets explicit: TestEval benchmarks test-case generation, and SWT-bench evaluates tests for validating real-world bug fixes~\cite{wang2025testeval,mundler2024swt}. These settings provide controlled tasks for assessing generated artifacts. We examine a different unit of analysis: the test-writing process within a multi-step repair trajectory. This process view captures whether an agent creates a new test, when it does so, whether it executes the file, and what assertion and printed-value patterns the file contains.

\paragraph{Testing During Repository-Level Issue Resolution}
SWE-bench established real-world GitHub issue resolution as an evaluation setting for language models~\cite{jimenez2024swebench}. Systems such as SWE-agent and AutoCodeRover study how models interact with repository environments and tools to carry out software-engineering tasks~\cite{yang2024sweagent,zhang2024autocoderover}. Recent systems span repository repair tools, autonomous coding agents, test-time-scaled repair, and role-based multi-agent development workflows~\cite{orwall2024moatless,devin2024,trae2025agent,qian2024chatdev}. Within these end-to-end workflows, testing is one activity among repository exploration, code editing, command execution, and patch validation. A final resolution score therefore does not show whether an agent wrote new tests or what resources accompanied that behavior. We study six LLMs under one lightweight, bash-only scaffold whose default prompt provides an advisory testing cue while leaving concrete testing decisions to the model. We then compare testing behavior, resolution outcomes, and resource use between baseline prompts and prompts that encourage or discourage new-test writing.

\paragraph{Software-Agent Trajectories}
Process-oriented studies complement final patch scores by examining the steps agents take during issue resolution. Bouzenia and Pradel analyze thought--action--result trajectories; Majgaonkar et al. compare successful and failed trajectories across code agents; Ceka et al. trace APR-agent workflows and identify test generation and regression-test selection as bottlenecks; and Chen et al. study errors across the software-development process rather than only the final code~\cite{bouzenia2025understanding,majgaonkar2025understanding,ceka2025understanding,chen2025beyond}. Systematic failure analysis also identifies diagnostic errors and unproductive loops in automated issue solving~\cite{liu2025failures}. Building on these process studies, including their attention to testing bottlenecks, we provide a cross-model account of file creation, timing, and execution under a lightweight scaffold. We further characterize assertion/print composition and compare paired prompt conditions in terms of both task outcomes and resource use.

%% file: sections/conclusion.tex
\section{Conclusion}
\label{sec:conclusion}

This paper studies how LLM agents write and execute tests during GitHub issue resolution, using 5{,}000 recorded runs from six LLMs under a lightweight scaffold. Models differ substantially in whether, when, and how they test. For every model in the content analysis, statements that print runtime values outnumber assertions that check conditions. Under prompts that encourage or discourage new test files, agents often start or stop writing them. None of the resolution differences from the original prompts is statistically significant, while API calls or token use differ substantially for three models. These test files combine automatic checks with code that exposes runtime values for inspection, connecting agent-written testing with program comprehension during repair. Identical resolution totals can also hide changes in which issues are solved, as the GPT-5.2 comparison shows. Evaluations should therefore report testing instructions, assertion and print content, test execution, task-level outcomes, and resource use alongside resolution rates.

%% file: sections/data_availability.tex
\section{Data Availability}
\label{sec:data_availability}

To support the reproduction of our study and facilitate future research, we provide an anonymized replication package containing the datasets and analysis scripts at \url{https://figshare.com/s/c72f69149860ea293f4f}.